\pdfoutput=1

\documentclass[12pt,a4paper]{article}

\usepackage{jheplike,ifpdf}

\hypersetup{pdftitle={},pdfcreator={},linkcolor=[rgb]{0.15,0.35,0.75},colorlinks=true,citecolor=[rgb]{0.675,0,0.2},urlcolor=[rgb]{0.15,0.35,0.65}}
\let\olditemize\itemize\renewcommand{\itemize}{\vspace{-2pt}\olditemize\setlength{\itemsep}{1pt}\setlength{\parskip}{0pt}\setlength{\parsep}{-0pt}}
\let\oldenumerate\enumerate\renewcommand{\enumerate}{\vspace{-4pt}\oldenumerate\setlength{\itemsep}{1pt}\setlength{\parskip}{0pt}\setlength{\parsep}{0pt}}

\newcommand{\eq}[1]{\vspace{-0.5pt}\begin{equation}#1\vspace{-0.5pt}\end{equation}}

\newcommand{\mi}{\raisebox{0.75pt}{\scalebox{0.75}{$\,-\,$}}}
\newcommand{\pl}{\raisebox{0.75pt}{\scalebox{0.75}{$\,+\,$}}}
\newcommand{\dz}{d\Omega}
\newcommand{\kn}{\Lambda_n(\alpha',k,z)}
\newcommand{\floor}[1]{\lfloor\hspace{-1pt}#1\hspace{-1pt}\rfloor}
\renewcommand{\phi}{\varphi}
\renewcommand{\tilde}{\widetilde}
\newcommand{\zexpl}[2]{(z_{#1}\hspace{-0.5pt}\mi z_{#2}\hspace{-0.5pt})}

\title{~\\[140pt]{\LARGE \mbox{String-Like Dual Models for Scalar Theories}}\\[-20pt]}
\author[1]{\vspace{-20pt}Christian~Baadsgaard}\affiliation[1]{Department of Physics, Princeton University, Princeton, NJ 08544, USA}\emailAdd{cjepsen@princeton.edu}
\author[2]{\!\!,\,N.~E.~J.~Bjerrum-Bohr}\author[2]{\!\!,\,Jacob Bourjaily}\author[2]{and Poul~H.~Damgaard}\affiliation[2]{Niels Bohr International Academy and Discovery Center, University of Copenhagen\\The Niels Bohr Institute, Blegdamsvej 17, DK-2100, Copenhagen \O, Denmark}\emailAdd{bjbohr@nbi.dk}\emailAdd{bourjaily@nbi.ku.dk}\emailAdd{phdamg@nbi.dk} 
\abstract{
We show that all tree-level amplitudes in $\phi^p$ scalar field theory can be represented as the $\alpha'\!\to\!0$ limit of an $SL(2,{\mathbb R})$-invariant, string-theory-like dual model integral. These dual models are constructed according to constraints that admit families of solutions. We derive these dual models, and give closed formulae for all tree-level amplitudes of any $\phi^p$ scalar field theory.
}

\begin{document}\maketitle
\vspace{-0pt}\section{Introduction}\label{sec:introduction}\vspace{-10pt}
String theory has been extremely useful in providing alternative avenues for representing amplitudes in quantum field theory. One reason for this is string theory's totally different organization of terms in the scattering amplitudes, and the manner in which it is completely detached from conventional Feynman diagram evaluations. Irrespective of string theory's potential as a unified description of nature, the formalism has provided us with novel tools with which field theory amplitudes can be understood. This was noted early on~\cite{Mangano} and the whole field of string-based rules for scattering amplitudes in field theory~\cite{BKstring} illustrates this. Other examples include the simple and unified proof of field theory identities such as Kleiss-Kuijf relations~\cite{KK} and BCJ relations~\cite{BCJ} in terms of monodromy in string theory~\cite{BjerrumBohr:2009rd,Stieberger:2009hq}. Much additional information about field theory amplitudes can indeed be obtained from string theory in this way~\cite{BjerrumBohr:2010zs,Mafra:2011kj}. Related ideas continue to provide new insight into amplitude calculations~\cite{Bjerrum-Bohr:2016juj}. Another striking case is the KLT relations~\cite{KLT} between graviton amplitudes and gauge field amplitudes, the field theory momentum kernel of which~\cite{BjerrumBohr:2010ta} follows immediately from the more general momentum kernel at the level of string theory~\cite{BjerrumBohr:2010hn}. The CHY formalism~\cite{Cachazo:2013gna,Cachazo:2013hca,Cachazo:2013iea} based on scattering equations provides another interesting example of how string theory, and modifications thereof~\cite{Mason:2013sva,Berkovits:2013xba,Adamo:2013tsa,Bjerrum-Bohr:2014qwa,Geyer:2015bja,Geyer:2016wjx}, can provide new insight into the computation of field theory amplitudes. These more recent examples suggest that there is still much more to gain from exploring the way string theory(-like) amplitudes can be computed, even if one is only interested in the field theory limit.
 
An obvious clue comes from duality. This was evident already from the beginning of string theory, where the duality of the Veneziano amplitude~\cite{Veneziano} shows the possibility of treating different scattering channels in a unified manner. A multitude of field theory diagrams can correspond to a single string theory diagram. The Koba-Nielsen formula~\cite{KobaNielsen} generalizes this to an $n$-particle scattering amplitude. From various directions, one is led towards an interpretation reminiscent of Feynman diagrams in field theory~\cite{Kikkawa:1969qy,Kikkawa:1970qz,Gross:1970rg,Sakita:1970ep,Nielsen:1970bc}. Indeed, in modern language this can be seen as an alternative way of generating $\phi^3$-theory amplitudes if one lets the Regge slope $\alpha'$ approach zero~\cite{Scherk}. This unusual way of producing tree-level amplitudes for scalar $\phi^3$-theory immediately raises the question of whether other types of scalar interactions can be generated in a similar way. We wish to answer that question here. 

The original motivation for the present study came from our derivation of how the CHY formalism of scattering equations could be understood in terms of Feynman diagrams at both tree level and loop level~\cite{Baadsgaard:2015voa,Baadsgaard:2015ifa,Baadsgaard:2015hia,Baadsgaard:2015twa}. If a suitable string-theory-like integration measure could be established for more general scalar field theories, the transcription between string integrands and CHY integrands~\cite{Bjerrum-Bohr:2014qwa} would then possibly provide the compact prescription for generating general scalar field theories based on scattering equations. As will be explained below, the situation is slightly more complicated, both from the perspective of string theory (or, more appropriately, generalized dual models) and scattering equations. 

A first step towards understanding scalar field theories beyond $\phi^3$-theory in the scattering equation formalism has been taken by Cachazo, He and Yuan~\cite{Cachazo:2014nsa} using an elegant dimensional reduction argument for Yang-Mills theory and making a corresponding judicious choice of polarization vectors that projects dimensionally reduced Yang-Mills gauge connections onto just one scalar degree of freedom. The quartic Yang-Mills vertex then yields the sought-for scalar $\phi^4$ interaction vertex.\footnote{\footnotesize {Suggestions for generating $\phi^4$-theory as a limit of string theory have been considered in~\cite{Frizzo:1999zx,Marotta:1999re}.}} At its simplest level, this procedure is therefore obviously limited to scalar interactions of $\phi^4$ type. To go beyond, one might again gain insight from string theory and consider the next terms in the $\alpha'$-expansion. This is not straightforward, because in order to use the map~\cite{Bjerrum-Bohr:2014qwa} between string theory integrands and CHY integrands these integrands must be manifestly tachyon-free, and the leading $\alpha'$-correction (which could potentially yield $\phi^6$-vertices) vanishes for the superstring. An alternative way of generating higher $\phi^p$-theories using the CHY measure was presented in \mbox{ref.~\cite{Baadsgaard:2015ifa}}. The essential mechanism there was the generation of clusters of vertices with any number of legs using basic $\phi^3$ vertices and corresponding summation over propagators. This is clearly a rather indirect prescription. Here, instead, we return to the question of generating such theories in the context of (generalized) dual models, leaving a potential description in terms of CHY integrands as an open problem.

\newpage
This paper is organized as follows. In \mbox{section~\ref{sec:review_and_summary}} we review how amplitudes in scalar $\phi^3$-theory can be obtained in the $\alpha'\!\to\!0$ limit of string theory, and present the generalized dual models we have found. The derivation of these new models will be described in \mbox{section~\ref{sec:derivations}}, where we will take care to describe their generalizations. We will conclude with some forward-looking remarks about the possible interpretation of these models in \mbox{section~\ref{sec:conclusions}}.

\vspace{-10pt}\section{String-Like, Dual Models for Scalar Field Theories}\label{sec:review_and_summary}\vspace{-8pt}
The string-like dual models we have found are natural generalizations of the way in which scalar amplitudes in scalar $\phi^3$-theory are represented in string theory in the $\alpha'\!\to\!0$ limit. Therefore, it will be useful to briefly review this well-known story. Color-ordered scattering amplitudes in scalar $\phi^3$ field theory arise in the $\alpha'\!\to\!0$ limit of string theory in following form:
\vspace{-2pt}\eq{\mathcal{A}_n^{\phi^3}\!=\lim_{\alpha'\to0} (\alpha')^{n-3}\!\int\!\!\dz\,\,\,\kn\,\mathcal{I}_n^{3}(z)\quad\text{with}\quad\mathcal{I}^3_n\equiv\prod_{i=1}^{n}\frac{1}{(z_i-z_{i+1})},\label{phi3_rep}\vspace{-5pt}}
where the auxiliary variables $z_i$ are cyclically ordered (with $z_{n+1}\!=\!z_1$ understood), $\Lambda$ denotes the Koba-Nielsen factor~\cite{KobaNielsen}, 
\vspace{-2pt}\eq{\kn\equiv\prod_{i<j}(z_i-z_j)^{\alpha's_{ij}}\qquad\text{where}\qquad s_{ij}\equiv (k_i+k_j)^2,\label{koba_nielsen_factor}\vspace{-5pt}}
and $\dz$ is the integration measure of string theory, which we may define as:
\vspace{-2pt}\eq{\dz\equiv\delta(z_a-z_a^0)\delta(z_b-z_b^0)\delta(z_c-z_c^0)\times(z_a-z_b)(z_b-z_c)(z_c-z_a)\prod_idz_i\,\theta(z_{i}-z_{i+1}),\label{definition_of_string_measure}\vspace{-5pt}}
where the Heaviside functions, denoted $\theta(z)$ above, encode the ranges of integration. As is well known from string theory, the formula (\ref{phi3_rep}) is $SL(2,\mathbb{R})$-invariant, which ensures that we may choose $z_a, z_b, z_c$ as well as their gauge-fixings, $z_a^0,z_b^0,z_c^0$, freely. 

In order to illustrate our new string-like formulae, it will be useful to define: 
\vspace{-5pt}\eq{\mathcal{P}_n^{j}(z)\equiv\prod_{i=1}^n{(z_i-z_{i+j})}.\label{general_factor}\vspace{-6pt}}
In terms of this, we can give a formula for any (connected) $n$-point amplitude in $\phi^p$-theory as,
\vspace{-1pt}\eq{\mathcal{A}_n^{\phi^p}\equiv\lim_{\alpha'\to0} (\alpha')^{\frac{(n-p)}{(p-2)}}\left(\gamma^{p,0}\right)^{\frac{(2-n)}{(p-2)}}\!\int\!\!\dz\,\,\,\kn\,\,\,\mathcal{I}_{n}^{p,0}(z),\label{general_x0_amplitude}\vspace{-2pt}}
in terms of the integrand $\mathcal{I}^{p,0}_n(z)$, defined according to,
\vspace{-2pt}\eq{\mathcal{I}_n^{p,0}(z)\equiv\frac{1}{\mathcal{P}_{n}^q}\!\!\prod_{j=1}^{\left\lfloor\hspace{-1pt}\frac{n-2}{p-2}/2\hspace{-1pt}\right\rfloor}\frac{\left(\mathcal{P}_n^{(p-2)j+1}\right)^2}{\mathcal{P}_n^{(p-2)j}\mathcal{P}_n^{(p-2)j+4-p}},\quad q\!\equiv\!\left\{\begin{array}{l@{$\;\;\;\;\;\;$}l}\floor{n/2}& \frac{(n-2)}{(p-2)}\!\in\!(2\mathbb{Z}),\\(n\mi p\pl4)/2&\text{else},\end{array}\right.\label{general_x0_integrand}\vspace{-6pt}}
and for which $\gamma^{p,0}$ is a momentum-independent constant, 
\vspace{-4pt}\eq{\gamma^{p,0}\equiv \!\int\!\!\dz\,\,\,\mathcal{I}_{p}^{p,0}(z)= \frac{\pi^{p-\frac{7}{2}}}{(p-2)}\frac{\Gamma\left(\frac{p-2}{2}\right)}{\Gamma\left(\frac{p-1}{2}\right)}\label{prefactor_constant_x0}\vspace{-4pt}\,.\vspace{-4pt}}
The first few values of this constant are:
\eq{
\gamma^{3,0}=\gamma^{4,0}=1, \hspace{25pt}
\gamma^{5,0}=\pi^2/6, \hspace{25pt}
\gamma^{6,0}=\pi^2/3, \hspace{25pt}
\gamma^{7,0}=3\pi^4/40\,.
}
As the reader may infer, we will later find it possible to generalize the expression (\ref{general_x0_amplitude}) to a family parameterized by $x$, with an integrand $\mathcal{I}_n^{q,x}(z)$ and constant $\gamma_n^{p,x}$.

It should be clear that the integrand in (\ref{general_x0_integrand}) is $SL(2,\mathbb{R})$-invariant, has the correct weights, and the power of $\alpha'$ provides the correct scaling dimensions. While not completely obvious, it is a simple exercise to see that when $p\!=\!3$, the representation (\ref{general_x0_amplitude}) matches the string-theory expression (\ref{phi3_rep}) exactly. 

It is worthwhile to illustrate this formula in a few particular instances. For example, the 6-particle amplitude in $\phi^4$-theory would be given by,
\vspace{-2pt}\eq{\mathcal{I}^{4,0}_6=\frac{\mathcal{P}^3_6}{\left(\mathcal{P}_6^2\right)^2}=\frac{\zexpl{1}{4}^2\zexpl{2}{5}^2\zexpl{3}{6}^2}{\zexpl{1}{3}^2\zexpl{2}{4}^2\zexpl{3}{5}^2\zexpl{4}{6}^2\zexpl{5}{1}^2\zexpl{6}{2}^2}\,,\vspace{-2pt}}
while the 10-particle amplitude would be given by,
\vspace{-2pt}\eq{\mathcal{I}^{4,0}_{10}=\frac{\left(\mathcal{P}^3_{10}\right)^2\mathcal{P}^5_{10}}{\left(\mathcal{P}_{10}^2\mathcal{P}_{10}^4\right)^2}=\prod_{i=1}^{10}\frac{\zexpl{i}{i+3}^2\zexpl{i}{i+5}}{\zexpl{i}{i+2}^2\zexpl{i}{i+4}^2}\,.\vspace{-2pt}}
In fact, $\phi^4$-theory is simple enough for us to write down a relatively compact expression for any multiplicity. Expanding the general expression (\ref{general_x0_integrand}), we find:
\eq{\begin{split}
\mathcal{I}_n^{4,0}&=\frac{1}{\mathcal{P}_n^{n/2}}\hspace{-10pt}\prod_{j=1}^{\floor{(n-1)/4}}\hspace{-2pt}\left(\frac{\mathcal{P}_{n}^{2j+1}}{\mathcal{P}_{n}^{2j}}\right)^2\,,\\&=\left\{\begin{array}{l@{$\;\;$}l}\displaystyle\prod_{i=1}^n\left(\frac{\zexpl{i}{i+3}^2\cdots\zexpl{i}{i+n/2-2}^2\zexpl{i}{i+n/2}}{\zexpl{i}{i+2}^2\zexpl{i}{i+4}^2\cdots\zexpl{i}{i+n/2-1}^2}\right)&\frac{n}{2}\!\in\!(2\mathbb{Z}\pl1)\,;\\
\displaystyle\prod_{i=1}^n\left(\frac{\zexpl{i}{i+3}^2\zexpl{i}{i+5}^2\cdots\zexpl{i}{i+n/2-1}^2}{\zexpl{i}{i+2}^2\cdots\zexpl{i}{i+n/2-2}^2\zexpl{i}{i+n/2}}\right)&\frac{n}{2}\!\in\!(2\mathbb{Z})\,.
\end{array}\right.\end{split}}

It is not hard to generate corresponding expressions from (\ref{general_x0_integrand}) for any particular case of interest. Just for the sake of illustration, let us give a few more concrete examples. The 8-point amplitude in $\phi^5$-theory would be given by,
\vspace{-2pt}\eq{\mathcal{I}^{5,0}_8=\frac{\mathcal{P}^4_8}{\mathcal{P}_8^2\mathcal{P}_8^3}=\prod_{i=1}^8\frac{\zexpl{i}{i+4}}{\zexpl{i}{i+2}\zexpl{i}{i+3}}\,;\vspace{-2pt}}
and the 11-point amplitude by the integrand,
\vspace{-2pt}\eq{\mathcal{I}^{5,0}_{11}=\frac{\left(\mathcal{P}^4_{11}\right)^2}{\mathcal{P}_{11}^2\mathcal{P}_{11}^3\mathcal{P}_{11}^5}=\prod_{i=1}^{11}\frac{\zexpl{i}{i+4}^2}{\zexpl{i}{i+2}\zexpl{i}{i+3}\zexpl{i}{i+5}}\,.\vspace{-2pt}}
Similarly, the 14-particle amplitude in $\phi^6$-theory would be generated by the following integrand,
\vspace{2pt}\eq{\mathcal{I}^{6,0}_{14}=\frac{(\mathcal{P}^5_{14})^2}{\mathcal{P}^2_{14}\mathcal{P}^4_{14}\mathcal{P}^6_{14}}=\prod_{i=1}^{14}\frac{\zexpl{i}{i+5}^2}{\zexpl{i}{i+2}\zexpl{i}{i+4}\zexpl{i}{i+6}}\,;\vspace{-2pt}}
and the 22-point amplitude in $\phi^7$-theory by
\vspace{-2pt}\eq{\mathcal{I}^{7,0}_{22}=\frac{(\mathcal{P}^6_{22})^2\mathcal{P}_{22}^{11}}{\mathcal{P}_{22}^{2}\mathcal{P}_{22}^{5}\mathcal{P}_{22}^{7}\mathcal{P}_{22}^{10}}=\prod_{i=1}^{22}\frac{\zexpl{i}{i+6}^2\zexpl{i}{i+11}}{\zexpl{i}{i+2}\zexpl{i}{i+5}\zexpl{i}{i+7}\zexpl{i}{i+10}}\,.\vspace{-2pt}}

In the next section, we derive the formula for $\mathcal{I}_n^{p,0}(z)$ in (\ref{general_x0_integrand}), discuss the origins of the constant prefactors $\gamma_n^{p,0}$, show how these dual models arise from basic principles, and can be generalized in several interesting ways.

\vspace{19pt}\section{Derivation and Generalizations}\label{sec:derivations}\vspace{-2pt}
In this section, we derive the formula (\ref{general_x0_integrand}), explain the meaning of the prefactor $\gamma_n^{p,0}$, and see how it can be naturally be generalized in a number of ways. Let us begin with some general (fairly trivial) considerations that will allow us to define some important notation, and see how the basic ingredients in (\ref{general_x0_integrand}) emerge.

\vspace{-2pt}\subsection{General Considerations and Notation}\label{subsec:notation_etc}\vspace{-2pt}
We seek to construct string-theory-like dual models that generalize the representation of $\phi^3$-theory according to (\ref{phi3_rep}). 
Recall that connected tree amplitudes in $\phi^p$-theory are only non-zero when \mbox{$n\!=\!L(p\mi2)\pl2$} for some integer $L\!\in\!\mathbb{Z}$, where $L\mi1$ provides the number of propagators of the amplitude. We are interested in expressing such non-zero amplitudes as integrals of the following form:
\eq{(\alpha')^{L-1}\,\gamma\!\int\!\!\dz\,\,\,\kn\,\,\,\mathcal{I}_n(z).\label{general_integrands_of_interest}}
We have allowed for a kinematic-independent prefactor $\gamma$ in part to emphasize that we are going to be careful about such things (up to an overall sign); but also because, as we will see later on, that the form of $\gamma$ required to reproduce scattering amplitudes precisely will turn out to be quite interesting. 

The possible integrands $\mathcal{I}(z)$ appearing in (\ref{general_integrands_of_interest}) should be required to be $SL(2,\mathbb{R})$-invariant. This implies that it must be constructed out of products of differences $\zexpl{i}{j}$,
\eq{\mathcal{I}_n(z)=\prod_{i<j}\zexpl{i}{j}^{c_{ij}},\label{overly_general_ints}}
for some numbers $c_{ij}$ (which we do not assume to be integers). Moreover, upon including the integration measure $\dz$, $SL(2,\mathbb{R})$-invariance requires that the weight of any $z_i$ must be $-2$:
\eq{\sum_{j\neq i}c_{ij}=-2\quad\forall i,\qquad\text{where}\quad c_{ij}\equiv c_{ji}\,.\label{conformal_weight_rule}}

In order for $\mathcal{I}(z)$'s constructed in this way to give rise to color-ordered amplitudes, it is necessary that the factors be cyclically-invariant. This requires that
\vspace{-2pt}\eq{c_{i,i+q}=c_{i-q,i}\qquad\forall\,i,q.\label{cyclicity}\vspace{-2pt}}
As a consequence we see that we may in fact define cyclic exponents,
\vspace{-2pt}\eq{e_j\equiv c_{i,i+j}.\vspace{-2pt}}
Notice that (\ref{cyclicity}) immediately implies that $e_j\!=\!e_{n-j}$ for all $j$. Because of this, we can always without loss of generality restrict our attention to $e_j$ with $j\!\leq\!\floor{n/2}$. 

Using the exponents $\{e_i\}$, we can rewrite the integrand (\ref{overly_general_ints}) as,
\vspace{-4pt}\eq{\mathcal{I}_n(z)\equiv\prod_{j=1}^{\floor{n/2}}\mathcal{P}_n^{e_j},\label{products_of_factors_form}\vspace{-4pt}}
where the constraint (\ref{conformal_weight_rule}) implies 
\vspace{-4pt}\eq{\sum_{j=1}^{n-1}e_j=-2. \label{SL2R_e_inv}\vspace{-4pt}}

In order to reproduce scattering amplitudes in the $\alpha'\!\to\!0$ limit of (\ref{general_integrands_of_interest}), it must be that divergences arise in order to cancel the vanishing power of $\alpha'$. The regions where the integrand develops divergences sufficient to contribute something non-vanishing in this limit are quite combinatorial in nature, and give rise to poles involving propagators. Thus, in order to reproduce scattering amplitudes, the exponents $e_j$ must be chosen carefully. Let us describe how this can be done presently. 

\vspace{-6pt}\subsection{Analysis of Divergences in the Limit $\alpha'\!\to\!0$}\label{subsec:divergences_etc}\vspace{-2pt}
The way string theory reproduces the correct (dimensionful) poles of field theory amplitudes in the $\alpha'\!\to\!0$ limit must necessarily be associated with a corresponding divergence in inverse powers of $\alpha'$ since only the dimensionless quantities $\alpha' s_{ij}$ appear in the initial expression. The prefactor of the integral provides the canceling powers of $\alpha'$. This means that we need to understand the rate at which the integral itself ({\it e.g.}, the integral without its prefactor) diverges in the $\alpha'\!\to\!0$ limit. It is this degree of divergence together with regions within the integration region where divergences occurs that will tell us how we can generate general scalar tree-level amplitudes in the $\alpha'\!\to\!0$ limit.

Because the measure $d\Omega$ enforces an ordering of the variables in (\ref{general_integrands_of_interest})---that is, $z_i\!<\!z_{i-1}$---a divergence in inverse powers of $\alpha'$ in the integral can only come about in regions of the domain where subsets of consecutive variables $z_i$, $z_{i+1}$, \ldots, $z_{i+m}$ tend to the same value. We can check whether an integral of the form (\ref{general_integrands_of_interest}) has such a divergence, by letting $\tau\!\equiv\!\{i,i\pl1,\ldots,i\pl m\}$, defining, $y_j\!\equiv\!z_i\mi z_j$ and $\tilde{y}_j\equiv y_j/y_{i+m}$ for $j\!\in\!\tau$, and then considering the $\epsilon\!\equiv\! y_{i+m}\!\to\!0$ region of the integral over $y_{i+m}$. In changing variables from $z_i$, $i\!\in\!\tau$, to variables $z_i,\tilde{y}_{i+1},\ldots,\tilde{y}_{i+m-1},\epsilon$:
\begin{itemize}
 \item from the measure $d\Omega$, we pick up a factor of $\epsilon^{m-1}$;
 \item from $\Lambda_n(\alpha',k,z)$, we pick up a factor of $\epsilon^{\alpha's_\tau}$ where $s_\tau\!\equiv\!\left(\sum_{j\in \tau}k_j\right)^2$;
\item from $\mathcal{I}_n(z)$, we pick up a factor of $\epsilon^{e_l}$ for each factor of $(z_j\mi z_{j+l})^{e_l}$ with \mbox{$j,j+l\!\in\!\tau$}. The variables $z_i$ to $z_{i+m}$ contain $m$ pairs of neighboring variables, $(m-1)$ pairs of next-nearest neighbors, and so forth, down to one pair of variables $m$ steps away. Consequently, the total factor we pick up from $\mathcal{I}_n(z)$ is $\epsilon^{\Sigma_m}$ where 
 \eq{
\Sigma_m \equiv m \,e_1+(m-1)e_2+\cdots+2e_{m-1}+e_m\,. 
 }
 \end{itemize}
In total then, the integral over $\epsilon$ reduces to:
\eq{
\int_0^{z_i}\!\!\!d\epsilon\;\; \epsilon^{m-1+\alpha's_\tau+\Sigma_m}\big(1+\mathcal{O}(\epsilon)\big) \,. \label{eint}
}
From this expression it is evident that there is a $1/\alpha'$ divergence when 
\eq{\Sigma_m = -m \label{divcrit}
\,,} in which case the integral evaluates to
\eq{
\frac{1}{\alpha'}\left(\frac{1}{s_\tau}+\mathcal{O}(\alpha')\right)\,.
}
So a divergence arising from variables $\{z_j| j\in\tau\}$, tending to the same value, results in the propagator carrying the external momenta $\{k_j| j\in\tau\}$.

The remaining integrals factor in two: integrals over $z_j$ with $j\!\!\notin\!\!\tau\backslash\{i\}$ and integrals over $\tilde{y}_j$ with $j\!\in\!\{i\pl1,i\pl2,\ldots,i\pl m\mi1\}$. Iteratively repeating the above reasoning to the remaining integrations, one can determine the overall degree of divergence in inverse powers of $\alpha'$. In general an integral of the form (\ref{general_integrands_of_interest}) can have several distinct divergent regions of the integration domain, and it is necessary to sum over all of them to get the leading term in $\alpha'$. We refer the reader to \mbox{ref.\ \cite{Baadsgaard:2015voa}} for more details.

\newpage
\vspace{-6pt}\subsection{Matching Divergences to Feynman Diagrams}\label{subsec:matching}\vspace{-2pt}
Turning our attention to $\phi^p$-theory, recall that the number of particles involved in any (connected) amplitude must be a multiple of $(p\mi2)$. In order to construct an integral expression for the full amplitude, there has to be a divergent region of the integration domain for each such possible propagator. In other words, equation (\ref{divcrit}) must be satisfied for 
$m$ equal to $(p\mi2)$, $2(p\mi2)$, \ldots, and $(L\mi1)(p\mi2)$. (Recall that $\!L\equiv\!(n\mi2)/(p\mi2)$.) At the same time, we must ensure that equation (\ref{divcrit}) is {\it not} satisfied for values of $m$ that are not divisible by $p\mi2$.
Given these conditions, the requirement  (\ref{SL2R_e_inv}) of $SL(2,\mathbb{R})$-invariance is equivalent to equation (\ref{divcrit}) with $m\!\equiv\!L(p\mi2)$. The conditions on the set $\{e_i\}$ can therefore be summarized by the following set of (not all independent) equations
\eq{
\Sigma_{l(p-2)}=-l(p-2), \text{ for } l\in\{1,2,\ldots,L\}. \label{econd}
}
For $\phi^3$-theory, these equations read
\eq{\begin{split}
-1&=e_1 \\
-2&=2e_1+e_2 \\
-3&=3e_1+2e_2+e_3 \label{s3n8}\\[-4pt]
 &\hspace{6.4pt}\vdots \\[-2pt]
n-2&=(n-2)e_1+(n-3)e_2+\cdots+e_{n-2},
\end{split}}
and have the unique solution $e_1\!=\!-1$ and $e_i\!=\!0$ for $i\!\neq\!1$. We recognize these as the standard dual model exponents (\ref{phi3_rep}) that lead to $\phi^3$-theory.

For $p\!>\!3$, the conditions (\ref{econd}) provide an under-determined set of equations. We can parametrize the solution space by introducing parameters $\!\{x_m\!\}$ and demanding 
\eq{\Sigma_m = -m + 1 + x_m}
when $m$ is not a multiple of $p-2$. As long as each $x_m$ is greater than minus one, the integral (\ref{eint}) converges, and we get no incorrect propagators.\footnote{One could also consider the case $x_m\!<\!-1$, in which case the integral (\ref{eint}), after analytically continuing, remains finite in the $\alpha'\!\to\!0$ limit. However, it will no longer be possible to take the $\alpha'\!\to\!0$ limit before evaluating the integral, and the residual integrations discussed in the next section will no longer yield momentum independent-factors. }

In summary, we impose the following conditions on the exponents $e_i$:
\eq{
\Sigma_i = \begin{cases}
-i & i=\!0\,\,\text{mod}\,(p\mi2)\,,\\
x_i-i+1 & \text{else}\,.
\end{cases} \label{finalconditions}
}
These equations are fairly straightforward to solve. If we adopt the convention that $\Sigma_i=0$ for $i\!<\!0$, then, for $\phi^4$-theory:
\eq{\begin{split}
e_i = \Sigma_i-2\Sigma_{i-1}+\Sigma_{i-2}= \begin{cases}
x_1 &i=1\\
-2-2x_{i-1} &i\in(2\mathbb{Z})\\
2+x_i+x_{i-2}  & \text{else},
\end{cases}\label{p4solutions}
\end{split}}%
and for $\phi^p$-theory with $p\!>\!4$:
\eq{\begin{split}
e_i= \Sigma_i-2\Sigma_{i-1}+\Sigma_{i-2}=\begin{cases}
x_1 & i=1\,\\
x_i-2x_{i-1}-1 &i=\!2\,\,\text{mod}\,(p-2)\,, \\
x_{i-2}-2x_{i-1}-1&i=\!0\,\,\text{mod}\,(p-2)\,,\\
x_i+x_{i-2}+2 &1<i=\!1\,\,\text{mod}\,(p-2)\,,\\
x_i-2x_{i-1}+x_{i-2}&\text{else}\,.
\end{cases}\label{pg4solutions}
\end{split}}

The solutions (\ref{p4solutions}) and (\ref{pg4solutions}) apply only to the first $\floor{n/2}$ exponents $e_i$, but as explained above, this suffices. The remaining exponents can be found from the relation $e_{n-i}\!=\!e_i$.

\vspace{-6pt}\subsection{Residual Integrations}\label{subsec:residual}\vspace{-2pt}
As long as the exponents $e_i$ satisfy (\ref{p4solutions}) or (\ref{pg4solutions}), the integral (\ref{general_integrands_of_interest}) will contain divergences corresponding to all the factorization channels of the $\phi^p$ tree-amplitude. But in order to identify  a full amplitude with the corresponding integral in the $\alpha'\!\to\!0$ limit, we must ensure that all Feynman diagrams obtained from that integral come dressed with the same numerical prefactor, which we can then cancel with the overall normalization factor $\gamma$. The prefactors arise due to the fact that after carrying out all the integrations that lead to divergences in inverse powers of $\alpha'$, following the reasoning of section~\ref{subsec:divergences_etc}, what remains is the product of $L$ residual integrals that are finite as $\alpha'$ tend to zero. To leading order in $\alpha'$, we can therefore set $\Lambda_n(\alpha',k,z)\!=\!1$ for those integrals so that all momentum-dependence disappears. 

Feynman diagrams related by cyclic interchanges of the external momenta will necessarily carry the same numerical prefactor. The same may not be true in general for Feynman diagrams of different topologies (corresponding to the polygon graphs in Table 1 of~\cite{Baadsgaard:2015ifa}). To ensure that the prefactors match, we must impose additional conditions on the exponents $e_i$.

Consider the factorization mentioned in the end of section~\ref{subsec:divergences_etc}. After performing the $\epsilon$-integral, the integral over $z_j$, $j\!\in\!\mathbb{Z}_n$, has factored into an integral over $z_j$, $j\!\in\!\{1,2,\ldots,i,i\pl m\pl1,\ldots,n\}$ and an integral over $\tilde{y}_j$, $j\!\in\!\{i\pl1,i\pl2,\ldots,i\pl m\mi1\}$. So in the $z$-integral, the variables $z_{i-1}$ and $z_{i+m+1}$ are now next-nearest neighbors. In order for the $z$- and $\tilde{y}$-integrals to match, we must therefore require that $e_2=e_{m+2}$. 
Because all the original variables $z_i$ to $z_{i+m}$ have all merged to the same value $z_i$, which is now the nearest-neighbor of $z_{i-1}$, we must also demand that $e_1 = \sum_{l=1}^{m+1}e_l$.

Extending the above considerations beyond nearest and next-nearest neighbors, we find that the full list of (not-independent) requirements of these two types can be stated thus: for any multiple $m$ of $(p\mi2)$ we require that:
\eq{
e_j=\begin{cases}\displaystyle\sum_{l=j}^{m+j}e_l & j=1,2,\ldots,p-3, \\
e_{j+m} & j=2,\ldots,p-3\,. \end{cases}
}
These conditions will be satisfied if we equate all the parameters $x_i$ in the expressions (\ref{p4solutions}) and (\ref{pg4solutions}) 
for the exponents. This being done, the $L$ residual integrations all evaluate to the same momentum-independent value. The overall normalization constant must therefore be chosen as:
\eq{
\gamma \equiv \left(\int d\Omega \,\,\,\mathcal{I}_p(z)\right)^{-L}\,.
}

\vspace{-2pt}\subsection{Generalized Dual Models for All Scalar Field Theories}\label{subsec:exempli_gratia}\vspace{-2pt}

We see that the requirement that all the correct poles arise results in a family of solutions for $e_i$, each of which have extra free parameters $x_i$. The requirement that all terms arise from integrations with identical coefficients implies that all these parameters must be identical: $x\!\equiv\!x_i$ for all $i$. Thus, we arrive at a one-parameter family of dual models for any scalar tree-amplitude with $p\!>\!3$.

For $\phi^4$-theory the exponents are given by,
\eq{e_j\equiv\left\{\begin{array}{l@{$\;\;\;\;\;\;\;$}l}x&j=1;\\-2(1\pl x)&j\!\in\!(2\mathbb{Z});\\2(1+x)&\text{else}.\end{array}\right.\label{general_exponent_solutions}}
When $p\!>\!4$ we have the following exponents:
\eq{e_j\equiv\left\{\begin{array}{l@{$\;\;\;\;\;\;\;$}l}x&j=1;\\-(1\pl x)&j=\hspace{2.4pt}\,0,2\,\,\text{mod}\,(p\mi2);\\2(1\pl x)&1\!<\!j=\!1\,\,\text{mod}\,(p\mi2);\\0&\text{else}.\end{array}\right.\label{general_exponent_solutions}}
By using these exponents together with equations (\ref{general_integrands_of_interest}) and (\ref{products_of_factors_form}), we arrive at the following generalized dual model for all $n$-point amplitudes in $\phi^p$ field theory:
\vspace{-1pt}\eq{\mathcal{A}_n^{\phi^p}\equiv\lim_{\alpha'\to0} (\alpha')^{L-1}\left(\gamma^{p,x}\right)^{-L}\!\int\!\!\dz\,\,\,\kn\,\,\,\mathcal{I}_{n}^{p,x}(z),\label{general_x_amplitude}\vspace{-2pt}}
where the generalized integrand $\mathcal{I}_n^{p,x}(z)$ is given by, 
\vspace{-2pt}\eq{\mathcal{I}_n^{p,x}(z)\equiv \left(\mathcal{P}^{1}_n\right)^x\left(\mathcal{I}_n^{p,0}\right)^{1+x}\,,\label{general_x_integrand}\vspace{-6pt}}
and the integration constants $\gamma_n^{p,x}$ are given by
\vspace{-4pt}\eq{\gamma_n^{p,x}\equiv \!\int\!\!\dz\,\,\,\mathcal{I}_{p}^{p,x}(z)\,.\label{prefactor_constant_x}\vspace{-4pt}}
%

\vspace{-6pt}\section{Conclusions and Discussion}\label{sec:conclusions}\vspace{-8pt}
The well-known relationship between string theory and cubic graphs in its field theory limit immediately leads to the question: can we construct scalar field theories based on quartic or quintic or higher-order vertices that similarly follow from one single integral representation? To deviate as little as possible from the original string theory setting we have here considered a scenario as similar to open string theory as possible. We have introduced an ordered set of $n$ real parameters $z_i$ integrated on the real line. We have insisted that the integrand be $SL(2,\mathbb{R})$-invariant, and we have defined the integration measure with the conventional Koba-Nielsen term times $SL(2,\mathbb{R})$-invariant factors that, if successful, should define for us the different kinds of scalar field theories in the $\alpha'\!\to\!0$ limit. 

We have found that cubic graphs play a special role in that the conditions that need to be fulfilled for the integral to generate $\phi^3$-theory in the $\alpha'\!\to\!0$ limit have a unique solution, the one already known. This sheds some new light on the observations of Scherk in the classic paper on dual models~\cite{Scherk}. Surprisingly, we have found that all other scalar field theories based on $\phi^p$ vertices can be generated as well. These theories require a bit more care, and the conditions needed to determine their integrands do not lead to unique solutions. Nevertheless, special solutions can be found in which, for given $n$ and given $p$, we can write down the corresponding $n$-point amplitude as the $\alpha'\!\to\!0$ limit of a single string-like integral. The succinct expressions automatically generate the sum over all color-ordered Feynman diagrams of $\phi^p$-theory for that $n$-point amplitude.

Our original motivation for this study was an aim towards establishing a CHY-type prescription for arbitrary scalar field theories. Let us therefore include some comments on this program. First, for $\phi^3$-theory the string-like construction is unique, and the integrand is just in the form for which the transcription between string theory integrands and CHY integrands is well established~\cite{Bjerrum-Bohr:2014qwa,Baadsgaard:2015voa}: ending up as the product of two `cycles'. Interestingly, even Yang-Mills theory can in CHY form be described entirely in terms of integrands composed of products of such cycles~\cite{Bjerrum-Bohr:2016axv}, again indicating the fundamental nature of cubic graphs underneath the formalism. Higher-order scalar field theories, as constructed in this paper, are not provided in that form. There are numerator factors that spoil an immediate transcription into CHY language. Since the specific case of $\phi^4$-theory, for which there does exist a CHY construction~\cite{Cachazo:2014nsa}, appears as intractable as any other $\phi^p$-theory with $p\!>\!4$ there is still hope that it may be possible to rewrite our integrands suitably (perhaps by partial fractioning) so as to end up with expressions that can transcribed to CHY form. This we leave as an open problem. A constructive solution can always be provided by brute-force expressions of arbitrary scalar graphs in terms of underlying cubic graphs and correspondingly canceling numerator factors that eliminate unwanted propagators. This program can be carried through systematically in the CHY formalism, as described in detail in \mbox{ref.\ \cite{Baadsgaard:2015ifa}}.

It seems unlikely that the integral constructions of this paper can be related to some sort of open string theory: how could they be produced by a world-sheet path integral on the disc with vertex operator insertions? This question is especially interesting in the context of {\it e.g.}\ \mbox{ref.\ \cite{Gaberdiel:1998fs}}. It is intriguing that dual models of this kind can be constructed so that tree-level amplitudes of arbitrary $\phi^p$-theories fall out in the $\alpha'\!\to\!0$ limit. What kind of deformation parameter could this $\alpha'$ correspond to? Only its `point-like' $\alpha'\!\to\!0$ limit plays any role here. We leave these questions for future work.

\section*{Acknowledgements}\vspace{-6pt}
We thank Paolo Di Vecchia and Peter Goddard for discussions and helpful suggestions. We would also like to thank Bo Feng for numerous enlightening discussions on related issues. This work was supported in part by the Danish National Research Foundation (DNRF91) and by a MOBILEX research grant from the Danish Council for Independent Research (JLB). 

\providecommand{\href}[2]{#2}\begingroup\raggedright\endgroup


\begin{thebibliography}{10}

\bibitem{Mangano}
M.~L. Mangano, S.~J. Parke, and Z.~Xu, ``{Duality and Multi-Gluon
  Scattering},''
\href{http://dx.doi.org/10.1016/0550-3213(88)90001-6}{{\em Nucl. Phys.} {\bf
  B298} (1988)  653--672}.

\bibitem{BKstring}
Z.~Bern and D.~A. Kosower, ``{A New Approach to One Loop Calculations in Gauge
  Theories},''
\href{http://dx.doi.org/10.1103/PhysRevD.38.1888}{{\em Phys. Rev.} {\bf D38}
  (1988)  1888}.

\bibitem{KK}
R.~Kleiss and H.~Kuijf, ``{Multi-Gluon Cross-Sections and Five Jet Production
  at Hadron Colliders},''
\href{http://dx.doi.org/10.1016/0550-3213(89)90574-9}{{\em Nucl. Phys.} {\bf
  B312} (1989)  616}.

\bibitem{BCJ}
Z.~Bern, J.~Carrasco, and H.~Johansson, ``{New Relations for Gauge-Theory
  Amplitudes},'' \href{http://dx.doi.org/10.1103/PhysRevD.78.085011}{{\em Phys.
  Rev.} {\bf D78} (2008)  085011},
\href{http://arxiv.org/abs/0805.3993}{{ arXiv:0805.3993 [hep-ph]}}.

\bibitem{BjerrumBohr:2009rd}
N.~Bjerrum-Bohr, P.~H. Damgaard, and P.~Vanhove, ``{Minimal Basis for Gauge
  Theory Amplitudes},''
  \href{http://dx.doi.org/10.1103/PhysRevLett.103.161602}{{\em Phys. Rev.
  Lett.} {\bf 103} (2009)  161602},
\href{http://arxiv.org/abs/0907.1425}{{ arXiv:0907.1425 [hep-th]}}.

\bibitem{Stieberger:2009hq}
S.~Stieberger, ``{Open \& Closed vs. Pure Open String Disk Amplitudes},''
\href{http://arxiv.org/abs/0907.2211}{{ arXiv:0907.2211 [hep-th]}}.

\bibitem{BjerrumBohr:2010zs}
N.~E.~J. Bjerrum-Bohr, P.~H. Damgaard, T.~Sondergaard, and P.~Vanhove,
  ``{Monodromy and Jacobi-Like Relations for Color-Ordered Amplitudes},''
  \href{http://dx.doi.org/10.1007/JHEP06(2010)003}{{\em JHEP} {\bf 06} (2010)
  003},
\href{http://arxiv.org/abs/1003.2403}{{ arXiv:1003.2403 [hep-th]}}.

\bibitem{Mafra:2011kj}
C.~R. Mafra, O.~Schlotterer, and S.~Stieberger, ``{Explicit BCJ Numerators from
  Pure Spinors},'' \href{http://dx.doi.org/10.1007/JHEP07(2011)092}{{\em JHEP}
  {\bf 07} (2011)  092},
\href{http://arxiv.org/abs/1104.5224}{{ arXiv:1104.5224 [hep-th]}}.

\bibitem{Bjerrum-Bohr:2016juj}
N.~E.~J. Bjerrum-Bohr, J.~L. Bourjaily, P.~H. Damgaard, and B.~Feng,
  ``{Analytic Representations of Yang-Mills Amplitudes},''
\href{http://arxiv.org/abs/1605.06501}{{ arXiv:1605.06501 [hep-th]}}.

\bibitem{KLT}
H.~Kawai, D.~Lewellen, and S.~Tye, ``{A Relation Between Tree Amplitudes of
  Closed and Open Strings},''
\href{http://dx.doi.org/10.1016/0550-3213(86)90362-7}{{\em Nucl. Phys.} {\bf
  B269} (1986)  1}.

\bibitem{BjerrumBohr:2010ta}
N.~Bjerrum-Bohr, P.~H. Damgaard, B.~Feng, and T.~Sondergaard, ``{Gravity and
  Yang-Mills Amplitude Relations},''
  \href{http://dx.doi.org/10.1103/PhysRevD.82.107702}{{\em Phys. Rev.} {\bf
  D82} (2010)  107702},
\href{http://arxiv.org/abs/1005.4367}{{ arXiv:1005.4367 [hep-th]}}.

\bibitem{BjerrumBohr:2010hn}
N.~Bjerrum-Bohr, P.~H. Damgaard, T.~Sondergaard, and P.~Vanhove, ``{The
  Momentum Kernel of Gauge and Gravity Theories},''
  \href{http://dx.doi.org/10.1007/JHEP01(2011)001}{{\em JHEP} {\bf 1101} (2011)
   001},
\href{http://arxiv.org/abs/1010.3933}{{ arXiv:1010.3933 [hep-th]}}.

\bibitem{Cachazo:2013gna}
F.~Cachazo, S.~He, and E.~Y. Yuan, ``{Scattering Equations and
  Kawai-Lewellen-Tye Orthogonality},''
  \href{http://dx.doi.org/10.1103/PhysRevD.90.065001}{{\em Phys. Rev.} {\bf
  D90} (2014) no. 6, 065001},
\href{http://arxiv.org/abs/1306.6575}{{ arXiv:1306.6575 [hep-th]}}.

\bibitem{Cachazo:2013hca}
F.~Cachazo, S.~He, and E.~Y. Yuan, ``{Scattering of Massless Particles in
  Arbitrary Dimensions},''
  \href{http://dx.doi.org/10.1103/PhysRevLett.113.171601}{{\em Phys. Rev.
  Lett.} {\bf 113} (2014) no. 17, 171601},
\href{http://arxiv.org/abs/1307.2199}{{ arXiv:1307.2199 [hep-th]}}.

\bibitem{Cachazo:2013iea}
F.~Cachazo, S.~He, and E.~Y. Yuan, ``{Scattering of Massless Particles:
  Scalars, Gluons and Gravitons},''
  \href{http://dx.doi.org/10.1007/JHEP07(2014)033}{{\em JHEP} {\bf 1407} (2014)
   033},
\href{http://arxiv.org/abs/1309.0885}{{ arXiv:1309.0885 [hep-th]}}.

\bibitem{Mason:2013sva}
L.~Mason and D.~Skinner, ``{Ambitwistor Strings and the Scattering
  Equations},'' \href{http://dx.doi.org/10.1007/JHEP07(2014)048}{{\em JHEP}
  {\bf 1407} (2014)  048},
\href{http://arxiv.org/abs/1311.2564}{{ arXiv:1311.2564 [hep-th]}}.

\bibitem{Berkovits:2013xba}
N.~Berkovits, ``{Infinite Tension Limit of the Pure Spinor Superstring},''
  \href{http://dx.doi.org/10.1007/JHEP03(2014)017}{{\em JHEP} {\bf 1403} (2014)
   017},
\href{http://arxiv.org/abs/1311.4156}{{ arXiv:1311.4156 [hep-th]}}.

\bibitem{Adamo:2013tsa}
T.~Adamo, E.~Casali, and D.~Skinner, ``{Ambitwistor Strings and the Scattering
  Equations at One Loop},''
  \href{http://dx.doi.org/10.1007/JHEP04(2014)104}{{\em JHEP} {\bf 1404} (2014)
   104},
\href{http://arxiv.org/abs/1312.3828}{{ arXiv:1312.3828 [hep-th]}}.

\bibitem{Bjerrum-Bohr:2014qwa}
N.~E.~J. Bjerrum-Bohr, P.~H. Damgaard, P.~Tourkine, and P.~Vanhove,
  ``{Scattering Equations and String Theory Amplitudes},''
  \href{http://dx.doi.org/10.1103/PhysRevD.90.106002}{{\em Phys. Rev.} {\bf
  D90} (2014) no. 10, 106002},
\href{http://arxiv.org/abs/1403.4553}{{ arXiv:1403.4553 [hep-th]}}.

\bibitem{Geyer:2015bja}
Y.~Geyer, L.~Mason, R.~Monteiro, and P.~Tourkine, ``{Loop Integrands for
  Scattering Amplitudes from the Riemann Sphere},''
  \href{http://dx.doi.org/10.1103/PhysRevLett.115.121603}{{\em Phys. Rev.
  Lett.} {\bf 115} (2015) no. 12, 121603},
\href{http://arxiv.org/abs/1507.00321}{{ arXiv:1507.00321 [hep-th]}}.

\bibitem{Geyer:2016wjx}
Y.~Geyer, L.~Mason, R.~Monteiro, and P.~Tourkine, ``{Two-Loop Scattering
  Amplitudes from the Riemann Sphere},''
\href{http://arxiv.org/abs/1607.08887}{{ arXiv:1607.08887 [hep-th]}}.

\bibitem{Veneziano}
G.~Veneziano, ``{Construction of a Crossing-Symmetric, Regge Behaved Amplitude
  for Linearly Rising Trajectories},''
\href{http://dx.doi.org/10.1007/BF02824451}{{\em Nuovo Cim.} {\bf A57} (1968)
  190--197}.

\bibitem{KobaNielsen}
Z.~Koba and H.~B. Nielsen, ``{Reaction Amplitude for $n$ Mesons: A
  Generalization of the Veneziano-Bardakci-Ruegg-Virasora Model},''
\href{http://dx.doi.org/10.1016/0550-3213(69)90331-9}{{\em Nucl. Phys.} {\bf
  B10} (1969)  633--655}.

\bibitem{Kikkawa:1969qy}
K.~Kikkawa, B.~Sakita, and M.~A. Virasoro, ``{Feynman-Like Diagrams Compatible
  with Duality I: Planar Diagrams},''
\href{http://dx.doi.org/10.1103/PhysRev.184.1701}{{\em Phys. Rev.} {\bf 184}
  (1969)  1701--1713}.

\bibitem{Kikkawa:1970qz}
K.~Kikkawa, B.~Sakita, M.~A. Virasoro, and S.~A. Klein, ``{Feynman-Like
  Diagrams Compatible with Duality II: General Discussion Including Nonplanar
  Diagrams},''
\href{http://dx.doi.org/10.1103/PhysRevD.1.3258}{{\em Phys. Rev.} {\bf D1}
  (1970)  3258--3266}.

\bibitem{Gross:1970rg}
D.~J. Gross, A.~Neveu, J.~Scherk, and J.~H. Schwarz, ``{The Primitive Graphs of
  Dual Resonance Models},''
\href{http://dx.doi.org/10.1016/0370-2693(70)90703-3}{{\em Phys. Lett.} {\bf
  B31} (1970)  592--594}.

\bibitem{Sakita:1970ep}
B.~Sakita and M.~A. Virasoro, ``{Dynamical Model of Dual Amplitudes},''
\href{http://dx.doi.org/10.1103/PhysRevLett.24.1146}{{\em Phys. Rev. Lett.}
  {\bf 24} (1970)  1146}.

\bibitem{Nielsen:1970bc}
H.~B. Nielsen and P.~Olesen, ``{A Parton View on Dual Amplitudes},''
\href{http://dx.doi.org/10.1016/0370-2693(70)90474-0}{{\em Phys. Lett.} {\bf
  B32} (1970)  203--206}.

\bibitem{Scherk}
J.~Scherk, ``{Zero-Slope Limit of the Dual Resonance Model},''
\href{http://dx.doi.org/10.1016/0550-3213(71)90227-6}{{\em Nucl. Phys.} {\bf
  B31} (1971)  222--234}.

\bibitem{Baadsgaard:2015voa}
C.~Baadsgaard, N.~E.~J. Bjerrum-Bohr, J.~L. Bourjaily, and P.~H. Damgaard,
  ``{Integration Rules for Scattering Equations},''
  \href{http://dx.doi.org/10.1007/JHEP09(2015)129}{{\em JHEP} {\bf 09} (2015)
  129},
\href{http://arxiv.org/abs/1506.06137}{{ arXiv:1506.06137 [hep-th]}}.

\bibitem{Baadsgaard:2015ifa}
C.~Baadsgaard, N.~E.~J. Bjerrum-Bohr, J.~L. Bourjaily, and P.~H. Damgaard,
  ``{Scattering Equations and Feynman Diagrams},''
  \href{http://dx.doi.org/10.1007/JHEP09(2015)136}{{\em JHEP} {\bf 09} (2015)
  136},
\href{http://arxiv.org/abs/1507.00997}{{ arXiv:1507.00997 [hep-th]}}.

\bibitem{Baadsgaard:2015hia}
C.~Baadsgaard, N.~E.~J. Bjerrum-Bohr, J.~L. Bourjaily, P.~H. Damgaard, and
  B.~Feng, ``{Integration Rules for Loop Scattering Equations},''
  \href{http://dx.doi.org/10.1007/JHEP11(2015)080}{{\em JHEP} {\bf 11} (2015)
  080},
\href{http://arxiv.org/abs/1508.03627}{{ arXiv:1508.03627 [hep-th]}}.

\bibitem{Baadsgaard:2015twa}
C.~Baadsgaard, N.~E.~J. Bjerrum-Bohr, J.~L. Bourjaily, S.~Caron-Huot, P.~H.
  Damgaard, and B.~Feng, ``{New Representations of the Perturbative
  $S$-Matrix},'' \href{http://dx.doi.org/10.1103/PhysRevLett.116.061601}{{\em
  Phys. Rev. Lett.} {\bf 116} (2016) no. 6, 061601},
\href{http://arxiv.org/abs/1509.02169}{{ arXiv:1509.02169 [hep-th]}}.

\bibitem{Cachazo:2014nsa}
F.~Cachazo, S.~He, and E.~Y. Yuan, ``{Einstein-Yang-Mills Scattering Amplitudes
  From Scattering Equations},''
  \href{http://dx.doi.org/10.1007/JHEP01(2015)121}{{\em JHEP} {\bf 1501} (2015)
   121},
\href{http://arxiv.org/abs/1409.8256}{{ arXiv:1409.8256 [hep-th]}}.

\bibitem{Frizzo:1999zx}
A.~Frizzo, L.~Magnea, and R.~Russo, ``{Scalar Field Theory Limits of Bosonic
  String Amplitudes},''
  \href{http://dx.doi.org/10.1016/S0550-3213(00)00200-5}{{\em Nucl. Phys.} {\bf
  B579} (2000)  379--410},
\href{http://arxiv.org/abs/hep-th/9912183}{{ arXiv:hep-th/9912183 [hep-th]}}.

\bibitem{Marotta:1999re}
R.~Marotta and F.~Pezzella, ``{Two Loop $\varphi^4$ Diagrams from String
  Theory},'' \href{http://dx.doi.org/10.1103/PhysRevD.61.106006}{{\em Phys.
  Rev.} {\bf D61} (2000)  106006},
\href{http://arxiv.org/abs/hep-th/9912158}{{ arXiv:hep-th/9912158 [hep-th]}}.

\bibitem{Bjerrum-Bohr:2016axv}
N.~E.~J. Bjerrum-Bohr, J.~L. Bourjaily, P.~H. Damgaard, and B.~Feng,
  ``{Manifesting Color-Kinematics Duality in the Scattering Equation
  Formalism},'' \href{http://dx.doi.org/10.1007/JHEP09(2016)094}{{\em JHEP}
  {\bf 09} (2016)  094},
\href{http://arxiv.org/abs/1608.00006}{{ arXiv:1608.00006 [hep-th]}}.

\bibitem{Gaberdiel:1998fs}
M.~R. Gaberdiel and P.~Goddard, ``{Axiomatic Conformal Field Theory},''
  \href{http://dx.doi.org/10.1007/s002200050031}{{\em Commun. Math. Phys.} {\bf
  209} (2000)  549--594},
\href{http://arxiv.org/abs/hep-th/9810019}{{ arXiv:hep-th/9810019 [hep-th]}}.

\end{thebibliography}
\end{document}